\documentclass[
reprint,
superscriptaddress,
amsmath,
amssymb,
aps,
longbibliography,
pra,
showpacs,
floatfix
]{revtex4-1}
\usepackage{hyperref}
\usepackage{graphicx}
\usepackage{xcolor} 

\usepackage{tabularx}
\usepackage{multirow}
\usepackage{dcolumn} 
\newcolumntype{d}[1]{D{.}{.}{#1}}
\newcolumntype{L}[1]{>{\raggedright\arraybackslash}p{#1}}
\newcolumntype{C}[1]{>{\centering\arraybackslash}p{#1}}
\newcolumntype{R}[1]{>{\raggedleft\arraybackslash}p{#1}}

\usepackage{braket}

\usepackage{comment}
\usepackage{enumitem}
\usepackage{soul}

\definecolor{color1}{rgb}{0,0.25,0.70}

\hypersetup{colorlinks=true, 
    linkcolor={color1},
    citecolor={color1}, 
    urlcolor={color1}
}

\begin{document}


\title{Giant effective magnetic fields from optically driven chiral phonons in 4$f$ paramagnets}

\author{Dominik~M.\ Juraschek}
\email{djuraschek@tauex.tau.ac.il}
\affiliation{School of Physics and Astronomy, Tel Aviv University, Tel Aviv 69978, Israel}
\affiliation{Harvard John A. Paulson School of Engineering and Applied Sciences, Harvard University, Cambridge, MA 02138, USA}

\author{Tom\'{a}\v{s}\ Neuman}
\affiliation{Harvard John A. Paulson School of Engineering and Applied Sciences, Harvard University, Cambridge, MA 02138, USA}

\author{Prineha Narang}
\email{prineha@seas.harvard.edu}
\affiliation{Harvard John A. Paulson School of Engineering and Applied Sciences, Harvard University, Cambridge, MA 02138, USA}

\date{\today}


\begin{abstract}
We present a mechanism by which optically driven chiral phonon modes in rare-earth trihalides generate giant effective magnetic fields acting on the paramagnetic $4f$ spins. With cerium trichloride (CeCl$_3$) as our example system, we calculate the coherent phonon dynamics in response to the excitation by an ultrashort terahertz pulse using a combination of phenomenological modeling and first-principles calculations. We find that effective magnetic fields of over 100 tesla can possibly be generated that polarize the spins for experimentally accessible pulse energies. The direction of induced magnetization can be reversed by changing the handedness of circular polarization of the laser pulse. The underlying process is a phonon analog of the inverse Faraday effect in optics that has been described recently, and which enables novel ways of achieving control over and switching of magnetic order at terahertz frequencies.
\end{abstract}

\maketitle


\section{Introduction}

Ultrashort laser pulses are able to change the magnetic order of materials within pico- or femtoseconds, orders of magnitude faster than conventional spin-based devices \cite{Kirilyuk2010,Nemec2018}. Usually, the electromagnetic field components of a laser pulse couple to electronic degrees of freedom of the magnetic ions, leading to the notion of ultrafast opto-magnetism \cite{Kampfrath2011,Kalashnikova2015,Tzschaschel2017,Kubacka2014,Schlauderer2019}. Recent studies have demonstrated that light can also couple to the spins indirectly by exciting coherent vibrations of the crystal lattice (phonons) that transfer angular momentum to the magnetic ions \cite{nova:2017,juraschek2:2017,Shin2018,Maehrlein2018,Juraschek2019,Juraschek2020_3,Juraschek2021,Disa2021,Mashkovich2021,Juraschek2021_5} or modulate the crystal structure into a transient state of modified magnetic order \cite{Radaelli2018,Gu2018,Khalsa2018,Fechner2018,Disa2020,Juraschek2020,Rodriguez-Vega2020,Afanasiev2021,Rodriguez-Vega2021,Stupakiewicz2021,Giorgianni2021}. These \textit{phono-magnetic} methods promise higher selectivity and lower dissipation than techniques based on opto-magnetic effects due to the lower energy of the excitation.

Particularly interesting are circularly polarized, or chiral, phonons, where the ions in a solid move on closed elliptical or circular orbits, generating angular momentum that can be transferred to the spins \cite{zhang:2014,Garanin2015,Nakane2018,Hamada2020,Ruckriegel2020,Streib2021}. Previously, nondegenerate chiral phonons have been described at the Brillouin-zone edges of materials with hexagonal symmetries \cite{Zhang:2015,Zhu2018,Gao2018,Chen2019,Zhang2020,Chen2021_propagatingchiral}, but their direct excitation with an ultrashort laser pulse is prohibited due to the large momentum mismatch between photons and phonons. In contrast, degenerate chiral phonons consist of superpositions of two orthogonal components of doubly or triply degenerate phonon modes that can be found at the Brillouin-zone center of materials with uniaxial or cubic symmetries and can therefore be resonantly excited with light. In recent years, a number of studies have shown or predicted effective magnetic fields arising from coherent chiral phonon driving that reach the milli tesla range \cite{nova:2017,juraschek2:2017,Juraschek2019,Geilhufe2021,Juraschek2020_3,Juraschek2021,Geilhufe2022}.

Here, we propose that optically driven chiral phonons in rare-earth trihalides produce giant effective magnetic fields that exceed those previously seen by several orders of magnitude. We predict, at the example of CeCl$_3$, that effective magnetic fields of over 100~tesla should be achievable, which polarize the paramagnetically disordered spins, for laser energies well within the damage threshold of the crystal. The mechanism allows for bidirectional control of the induced magnetization through phonon chirality that in turn can be controlled by the polarization of the laser pulse.


\section{Properties of cerium trichloride}


\begin{figure}[b]
\centering
\includegraphics[scale=0.077]{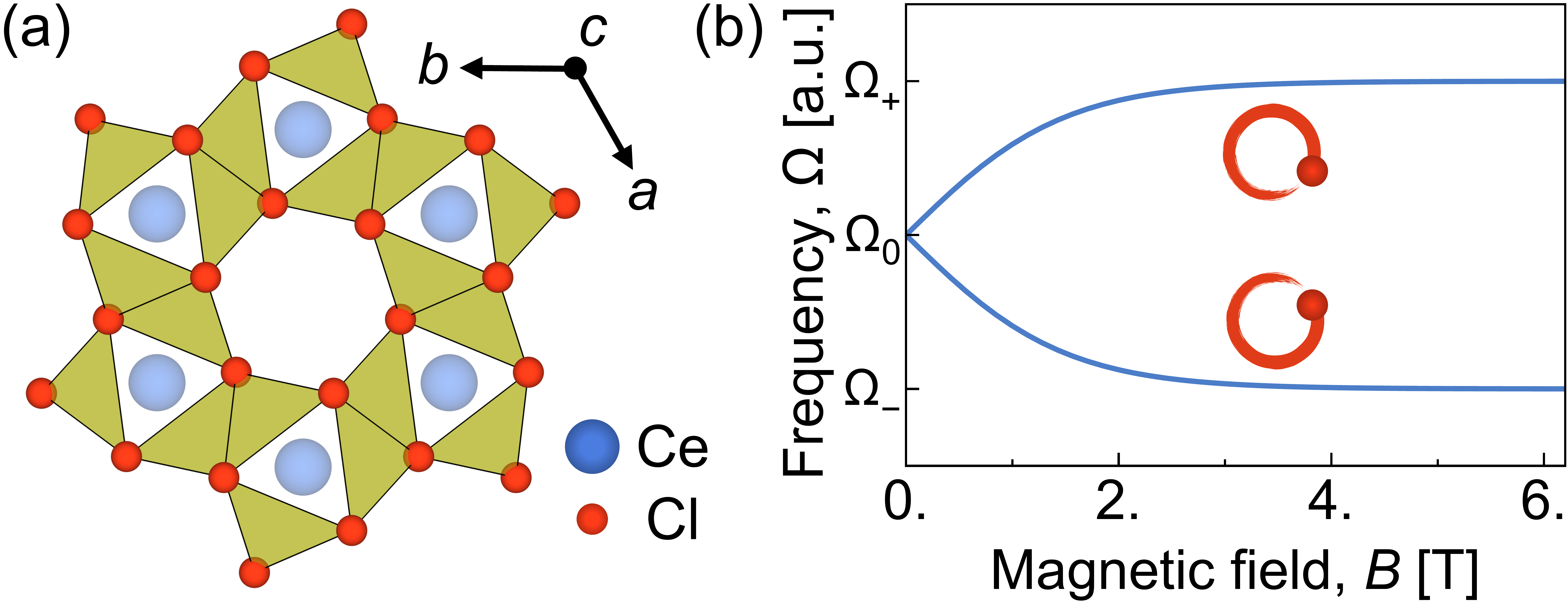}
\caption{
Structure and properties of CeCl$_3$. (a) Hexagonal $P6_3/m$ structure of paramagnetic CeCl$_3$. (b) Schematic splitting of a doubly degenerate phonon mode with frequency $\Omega_0$ into right- and left-handed circularly polarized (chiral) components in an external magnetic field, saturating at frequencies $\Omega_+$ and $\Omega_-$. At higher temperatures, the phonon splitting saturates at higher magnetic fields.
}
\label{fig:CeCl3}
\end{figure}

Rare-earth trihalides are a class of $4f$ paramagnets with formula unit $RH_3$. CeCl$_3$ ($R=\mathrm{Ce}$, $H=\mathrm{Cl}$) is a representative of this class of materials that crystallizes in the hexagonal $P6_3/m$ structure \cite{Zachariasen1948} with an electronic band gap of 4.2~eV \cite{Park1993}. Its Ce($4f^1$) valence-electron configuration remains paramagnetic for all temperature ranges relevant here, as spin ordering only occurs at very small temperatures of $<0.1$~K \cite{Landau1973}. We chose CeCl$_3$ as our model system, because the primitive unit cell consists of only 8 atoms (Fig.~\ref{fig:CeCl3}(a)), resulting in a small number of 21 optical phonon modes characterized by the irreducible representations $2A_g + 1A_u + 2B_g + 2B_u + 1E_{1g} + 3E_{2g} + 2E_{1u} + 1E_{2u}$ in its $6/m$ point group. Early Raman studies have shown that the polarization of the $4f$ electrons in an external magnetic field leads to a splitting of the doubly degenerate $E_{1g}$ and $E_{2g}$ phonon modes into left- and right-handed circular polarization \cite{schaack:1976,schaack:1977}, therefore obtaining chirality, see Fig.~\ref{fig:CeCl3}(b). It has been suggested that also the infrared-active $E_{1u}$ phonon modes split in the same way \cite{Thalmeier1978}, yet no experimental infrared spectroscopy measurements had been performed at that time. The infrared-active $E_{1u}$ modes map into the same $E'$ representation at the local $\bar{6}$ symmetry of the cerium ions as the Raman-active $E_{2g}$ modes, for which phonon splittings have been measured, and should therefore have the same effect on the paramagnetic spins. Infrared-active phonon modes possess an electric dipole moment and can therefore be resonantly excited by the electric field component of a laser pulse to yield large vibrational amplitudes. We will explore in this work how optically driven chiral $E_{1u}$ phonons act on the spins through the inverse of the spin-phonon coupling.


\section{Spin-phonon coupling and coherent phonon dynamics}

\begin{figure}[t]
\centering
\includegraphics[scale=0.115]{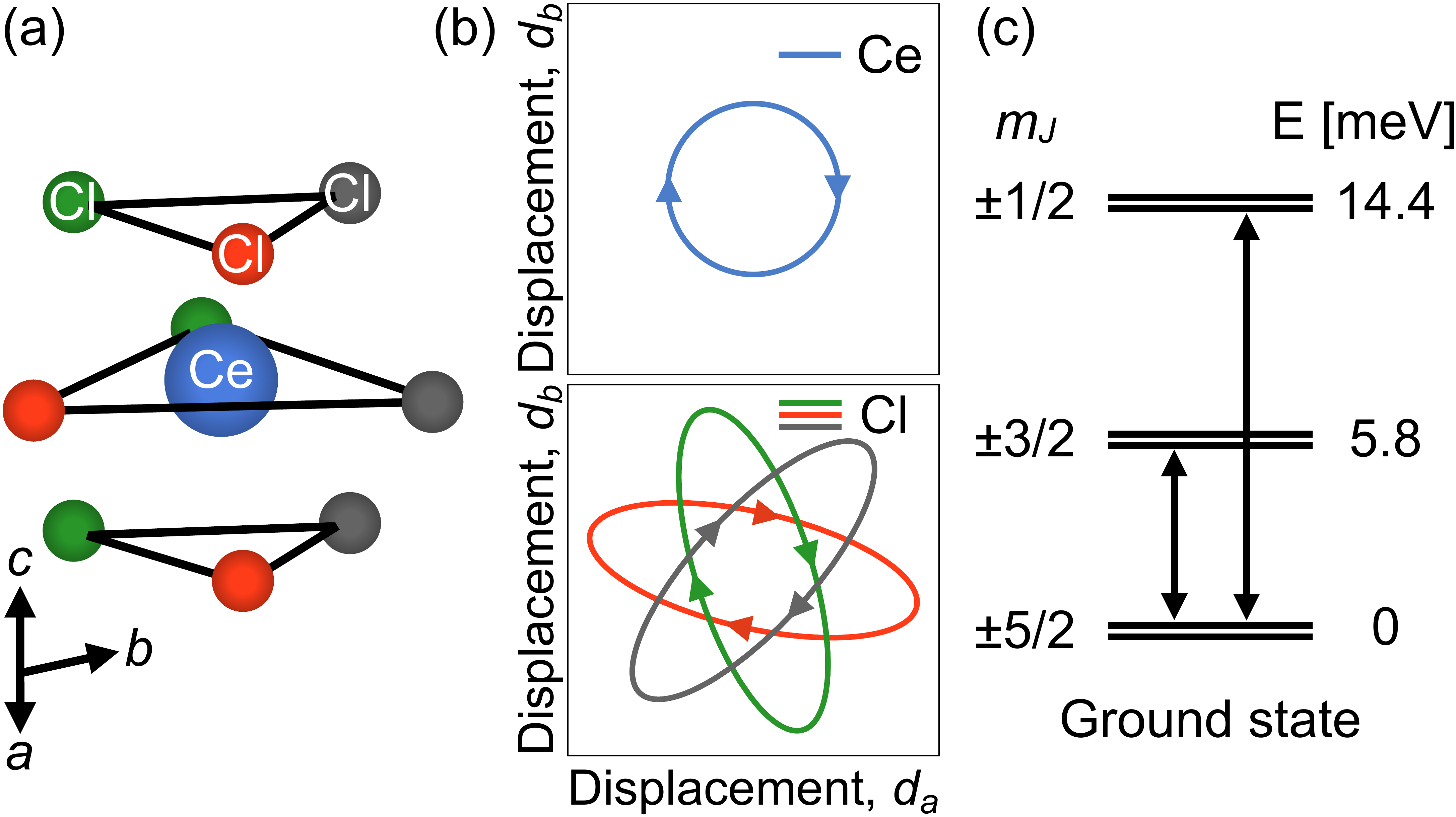}
\caption{
Spin-phonon coupling. (a) Cl ligands around the magnetic cerium ion. (b) Displacements of the Ce (left) and Cl ions (right) along the eigenvectors of the chiral $E_{1u}$ modes in the $ab$ plane of the crystal. The equilibrium positions of the ions are set to the center of each plot, respectively. (c) Circularly polarized phonons induce transitions between the $m_J=5/2$ ground-state Kramers doublet and higher crystal electric field levels \cite{schaack:1977,Thalmeier1977}. 
}
\label{fig:crystalfield}
\end{figure}

We begin by reviewing the theory of spin-phonon coupling in $4f$ paramagnets. Motions of the ions along the eigenvectors of doubly degenerate phonon modes modify the crystal electric field around the paramagnetic ions and induce virtual transitions between the ground-state energy levels and higher-lying states, see Fig.~\ref{fig:crystalfield}. The spin states in $4f$ paramagnets are close to those of the free ions and the total angular momentum (isospin), $J$, is a good quantum number. In CeCl$_3$, the lowest energy level has $J=5/2$, which splits into three Kramers doublets, of which $m_J=\pm5/2$ is the ground state. the interaction of chiral phonons with the isospin can be written as an effective ``spin-orbit'' type Hamiltonian \cite{Ray1967,Capellmann1991,Ioselevich1995,sheng:2006,Kagan2008,Wang2009,zhang:2014} 
\begin{equation}\label{eq:newspinphonon}
H^{\mathrm{s-ph}} =  K \mathbf{m} \cdot \mathbf{L},
\end{equation}
where $K$ is the coupling coefficient and $\mathbf{L}=\mathbf{Q}\times\dot{\mathbf{Q}}$ is the phonon angular momentum, with $\mathbf{Q}=(Q_{a},Q_{b},0)$ containing the normal mode coordinates of the two orthogonal components of a doubly degenerate phonon mode, $Q_{a}$ and $Q_{b}$, in the $ab$ plane of the crystal. $\mathbf{m}$ is the magnetic moment per unit cell, which, for components perpendicular to the $ab$ phonon polarizations, is given by
\begin{equation}\label{eq:magnetization}
\mathbf{m} = 2 g_J \mu_B \sqrt{J(J+1)} \mathbf{e}_z \left(\Braket{n_{-J}}-\Braket{n_J}\right),
\end{equation}
where $g_J$ is the Land\'{e} factor, $\mathbf{e}_z$ is a unit vector along the $c$ axis of the crystal and $\Braket{n_{\pm J}}$ are Fermi-Dirac distributions describing the occupation of the ground-state doublet. The theoretical value of the prefactor in Eq.~(\ref{eq:magnetization}) for the $m_J=\pm5/2$ ground-state doublet, $g_{\pm5/2} = g_{J} \sqrt{J(J+1)} = 2.54$, is reasonably close to the experimental value of 2.02 \cite{Thalmeier1977}, showing that most of the orbital angular momentum is unquenched. Please see the Appendix for detailed derivations.


We now look at the influence of the interaction on the phonons. The spin-phonon coupling modifies the off-diagonal terms of the dynamical matrix as $\mathbf{D}(m) = \mathbf{D}^{(0)} + i\mathbf{D}^{(1)}(m)$, where $m=|\mathbf{m}|$ \cite{anastassakis:1972,Holz1972,schaack:1976,schaack:1977,dzyaloshinskii:2009,riseborough:2010,juraschek2:2017,Liu2017_phonondichroism,Juraschek2020_3,Baydin2022}. 
As a result, the frequencies of right- and left-handed circular polarizations, $\Omega_\pm$, of the doubly degenerate phonon mode split,
\begin{equation}\label{eq:phononsplittinglinear}
\Omega_\pm(m) = \Omega_0\sqrt{1\pm\frac{2 K m}{\Omega_0}} \approx \Omega_0 \pm K m,
\end{equation}
where $\Omega_0$ is the eigenfrequency of the doubly degenerate phonon mode. Without an external magnetic field, the energy levels of the ground-state doublet are degenerate, there is no net magnetic moment per unit cell, and the phonon frequencies in Eq.~(\ref{eq:phononsplittinglinear}) remain degenerate. Applying a magnetic field, $B\parallel c$, to the paramagnet splits the ground-state doublet, $\Delta E = E_{-5/2}-E_{5/2} = 2 g_{\pm5/2} \mu_B B$, and, using Eq.~(\ref{eq:magnetization}), induces a splitting of the form 
\begin{equation}\label{eq:phononsplittingfull}
\Delta\Omega(B) = 2Km = 4 K g_{\pm5/2}\mu_B \tanh\left(\frac{g_{\pm5/2} \mu_B B}{2k_B T}\right).
\end{equation}
The prefactor in Eq.~(\ref{eq:phononsplittingfull}) directly corresponds to the saturation splitting, $\Delta\Omega_s=4 K g_{\pm5/2}\mu_B$, and we can reciprocally extract the spin-phonon coupling from experimentally measured phonon splittings, $K=\Delta\Omega_s/(4 g_{\pm5/2}\mu_B)$.


\begin{figure*}[t]
\centering
\includegraphics[scale=0.115]{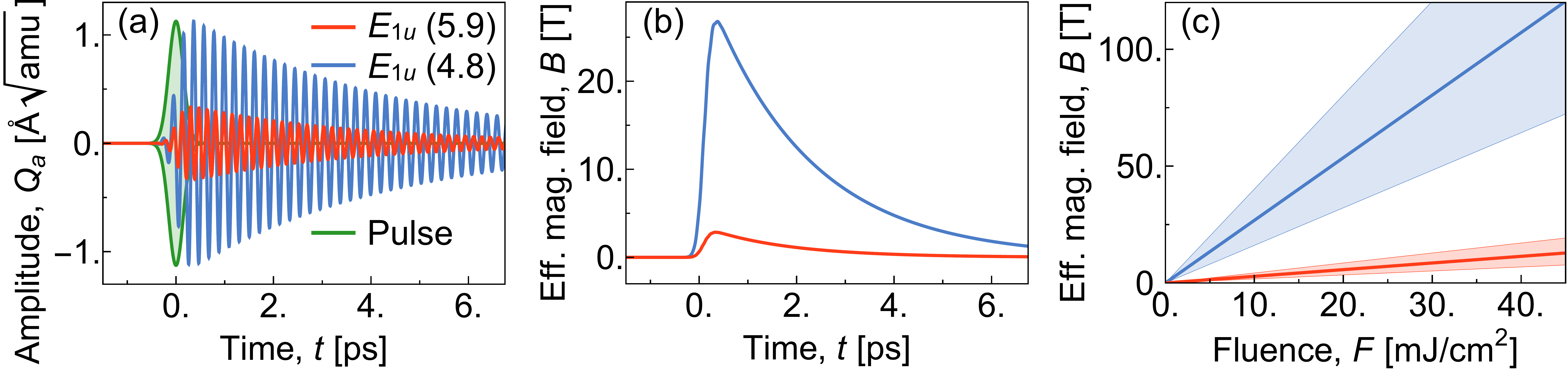}
\caption{
Coherent phonon dynamics and effective magnetic fields. (a) Time evolutions of the infrared-active 5.9 and 4.8~THz $E_{1u}$ phonon amplitudes, $Q_a$, in response to the excitation by a circularly polarized terahertz pulse with a full width at half maximum duration of $\tau=350$~fs and a fluence of 10~mJ/cm$^2$. The phases of the $Q_b$ components (not shown) are shifted by a quarter period, respectively. The carrier envelope of the terahertz pulse is shown schematically. (b) Time evolutions of the phonon-induced effective magnetic fields, $B$, acting on the paramagnetic spins. (c) Linear scaling of the effective magnetic fields with the fluence, $F$, of the terahertz pulse. The shaded area marks the range of magnetic fields that can be achieved for a range of commonly found spin-phonon coupling strengths.
}
\label{fig:dynamics}
\end{figure*}

We now look at the inverse effect the interaction has on the magnetization, when phonons of one type of chirality are driven with an ultrashort laser pulse. The phonon angular momentum acts as an effective magnetic field, $\mathbf{B}$,
\begin{equation}\label{eq:magneticfield}
\mathbf{B} = \partial H^\mathrm{s-ph} / (\partial \mathbf{m}) = K \mathbf{L}.
\end{equation}
Phenomenologically, this type of interaction has recently been described as a phonon analog of the inverse Faraday effect in optics \cite{Juraschek2020_3}, which is known to induce magnetizations in paramagnets \cite{vanderziel:1965,Pershan1967,reid:2010,Mikhaylovskiy2012}. A first experiment demonstrating this effect with elliptically polarized phonons has induced spin and magnetization dynamics in a complex oxide in recent years \cite{nova:2017}. If the spins reacted instantaneously to the effective magnetic field, the magnetization could be described statically as in Eq.~(\ref{eq:phononsplittingfull}). Experiments on the optical inverse Faraday effect have shown that this static limit of spin response holds for driving pulses on the order of nanoseconds \cite{vanderziel:1965,Pershan1967}. For femtosecond pulse durations, additional diamagnetic effects from the unquenching of electronic orbital moments come into play \cite{reid:2010,popova:2011,Popova2012,Mikhaylovskiy2012}, which cannot be described in the thermodynamic limit. Coherently driven phonons evolve over several picoseconds, which is also the timescale that spins and phonons have been shown to equilibrate through effective magnetic fields \cite{Maehrlein2018}. We therefore apply a rate-equation model to describe the dynamics of the spin population of the $m_J=\pm5/2$ ground-state doublet, $n_{\pm J}$ \cite{Breuer2003, Blum:1433745},
\begin{eqnarray}
\partial_t n_{\pm J} & = & -\gamma_{\pm J}(\Delta E) n_{\pm J} + \gamma_{\mp J}(\Delta E) n_{\mp J}, \label{eq:rateequation1}
\end{eqnarray}
where the decay rates of spins in the respective states are described by $\gamma_{-J} = \eta_0\Delta E  N(\Delta E)$ and $\gamma_{J} = \eta_0\Delta E (N(\Delta E)+1) $, where $N(\Delta E)$ is the Bose-Einstein distribution, $\eta_0=\gamma_0/(k_B T)$, and $\gamma_0$ is the decay rate for zero level splitting, $\gamma_{\pm J}(\Delta E \rightarrow 0)=\gamma_0$.

An ultrashort terahertz pulse can resonantly excite infrared-active phonons into a coherent quantum state, which allows us to treat the normal mode coordinate, $\mathbf{Q}$, as semi-classical field amplitude \cite{merlin:1997,Dekorsy2000,Forst2008,subedi:2014,fechner:2016,Juraschek2018}. We obtain $\mathbf{Q}$ by solving the equation of motion
\begin{equation}\label{eq:phononeom}
\ddot{\mathbf{Q}} + 2\kappa\dot{\mathbf{Q}} + \Omega^2_0\mathbf{Q} = Z \mathbf{E}(t).
\end{equation}
Here, $\kappa$ is the linewidth of the phonon mode, $\Omega_0$ is its eigenfrequency, and $Z=\sum_m Z^\ast_m \mathbf{q}_{m}/\sqrt{\mathcal{M}_m}$ is its mode effective charge, where $Z^\ast_m$ is the Born effective charge tensor, $\mathbf{q}_{m}$ is the eigenvector, and $\mathcal{M}_m$ is the atomic mass of ion $m$. The sum runs over all ions in the unit cell. We model the circularly polarized terahertz pulse as $\mathbf{E}(t)=(E(t),E(t-2\pi/(4\Omega)),0)/\sqrt{2}$, where $E(t) = E_0 \exp(-t^2/(2(\tau/\sqrt{8\ln2})^2)) \cos(\omega_0 t)$, $E_0$ is the peak electric field, $\omega_0$ is the center frequency, and $\tau$ is the full width at half maximum duration of the pulse. Here, the two perpendicular components of the doubly degenerate phonon mode are excited with a quarter-period difference, resulting in circular polarization and therefore chirality. As light couples to phonon modes close to the center of the Brillouin zone, we may neglect any wavevector dependence in Eq.~(\ref{eq:phononeom}).


\section{Computational details}

We calculate the phonon eigenfrequencies and eigenvectors, and Born effective charges from first principles, using the density functional perturbation theory formalism \cite{Gonze1997,Gonze1997_2} as implemented in the Vienna ab-initio simulation package (\textsc{vasp}) \cite{kresse:1996,kresse2:1996} and the frozen-phonon method as implemented in the \textsc{phonopy} package \cite{phonopy}. We use the VASP projector augmented wave (PAW) pseudopotentials with valence electron configurations Ce ($6s^2 5s^2 5p^6 5d^1 4f^1$) and Cl ($3p^5 3s^2$) and converge the Hellmann-Feynman forces to 25~$\mu$eV/\AA. For the 8-atom unit cell, we use a plane-wave energy cut-off of 600~eV, and a 4$\times$4$\times$7 gamma-centered $k$-point mesh to sample the Brillouin zone. For the exchange-correlation functional, we choose the Perdew-Burke-Ernzerhof revised for solids (PBEsol) form of the generalized gradient approximation (GGA) \cite{csonka:2009}. We perform nonmagnetic calculations to obtain the structural and dynamical properties of CeCl$_3$. A fully ab-initio treatment of paramagnetism in CeCl$_3$ would require supercell calculations and a description of $4f$ electron magnetism, which would make the computation of dynamical properties intractable for our purposes.
Within the nonmagnetic treatment, the lattice constants of our fully relaxed hexagonal structure (space group $P6_3/m$, point group $6/m$) of $a=4.21$~\AA{} and $c=7.38$~\AA{} with a unit-cell volume of $V_c=199$~\AA$^3$ agree reasonably well with experimental values \cite{Zachariasen1948}. Furthermore, our calculated phonon eigenfrequencies match the experimental values reasonably well \cite{schaack:1977,Thalmeier1978}, with a maximum deviation of $\sim$10\%{}. Crystal structures are visualized using \textsc{vesta} \cite{Momma2011}.


\section{Phonon-induced effective magnetic fields and magnetizations}

We extract the magnitude of the spin-phonon coupling from experimental data of the phonon-frequency splitting according to Eq.~(\ref{eq:phononsplittingfull}), $K=\Delta\Omega_s/(4g_{\pm5/2}\mu_B)$. In rare-earth trihalides, splittings of the Raman-active modes range between 0.3~THz and 0.75~THz \cite{schaack:1976,schaack:1977}. Because the infrared-active modes change the local symmetry of the magnetic cerium ion in the same way, we expect a similar strength of the spin-phonon coupling as for the Raman-active modes and use an average of the experimentally found values of $\Delta\Omega_s/(2\pi) = 0.5$~THz. This splitting is several orders of magnitude larger than the one induced by the magnetic moments of phonons in the phonon Zeeman effect \cite{Rebane1983,juraschek2:2017,Juraschek2019,Dunnett2019}, which we can therefore neglect here. Note that there are further microscopic origins of phonon angular momentum and magnetic moments, such as topological band features in semimetals \cite{Sengupta2020,Cheng2020,Xiao2021,Ren2021,Saparov2022,Li2021_topochiralphonon} or thermal gradients \cite{Hamada2018,Hamada2020_rotoelectric}, which however do not play a role here either.

In the following, we evaluate the effective magnetic fields produced by the two doubly degenerate infrared-active $E_{1u}$ modes in CeCl$_3$ with eigenfrequencies of 5.9 and 4.8~THz. We find the mode effective charges of these modes to be $0.24e$ and $0.66e$, respectively, where $e$ is the elementary charge. 
For the phonon linewidth, $\kappa$, we assume a phenomenological value of 5\%{} of the phonon frequency that matches those typically found in rare-earth trihalides \cite{schaack:1976,schaack:1977}. Fig.~\ref{fig:dynamics} shows the coherent phonon dynamics following the excitation by a circularly polarized terahertz pulse with a duration of $\tau=350$~fs and a fluence of 10~mJ/cm$^2$, as described by Eq.~(\ref{eq:phononeom}). The fluence $F$ is connected to the peak electric field and the duration of the pulse through $F=\tau/\sqrt{8\ln 2}c_0 \epsilon_0\sqrt{\pi/2}E_0^2$, where $c_0$ and $\epsilon_0$ are the speed of light and the vacuum permittivity. The center frequency, $\omega_0$, is chosen to be resonant with the eigenfrequencies of the respective phonon modes. In Fig.~\ref{fig:dynamics}(a), we show the evolution of the phonon amplitudes $Q_a$ according to Eq.~(\ref{eq:phononeom}). The phases of the $Q_b$ components are shifted by a quarter period, respectively. The maximum amplitude of the $E_{1u}(5.9)$ mode of $Q_a = 0.33$~\AA$\sqrt{\mathrm{amu}}$, where amu denotes the atomic mass unit, is roughly three times smaller than that of the $E_{1u}(4.8)$ mode of $Q_a = 1.1$~\AA$\sqrt{\mathrm{amu}}$ due to the smaller mode effective charge and higher phonon frequency. In Fig.~\ref{fig:dynamics}(b), we show the evolutions of the effective magnetic fields produced by the two chiral phonon modes according to Eq.~(\ref{eq:magneticfield}). We obtain a maximum effective magnetic field of $B=2.9$~T for the $E_{1u}(5.9)$ mode and $27$~T for the $E_{1u}(4.8)$ mode. This order-of-magnitude difference comes from the quadratic scaling of the effective magnetic field with the phonon amplitudes. The direction of the effective magnetic field is determined by the handedness of the phonon chirality, which can straightforwardly be controlled by the handedness of circular polarization of the pulse.

\begin{figure}[t]
\centering
\includegraphics[scale=0.0825]{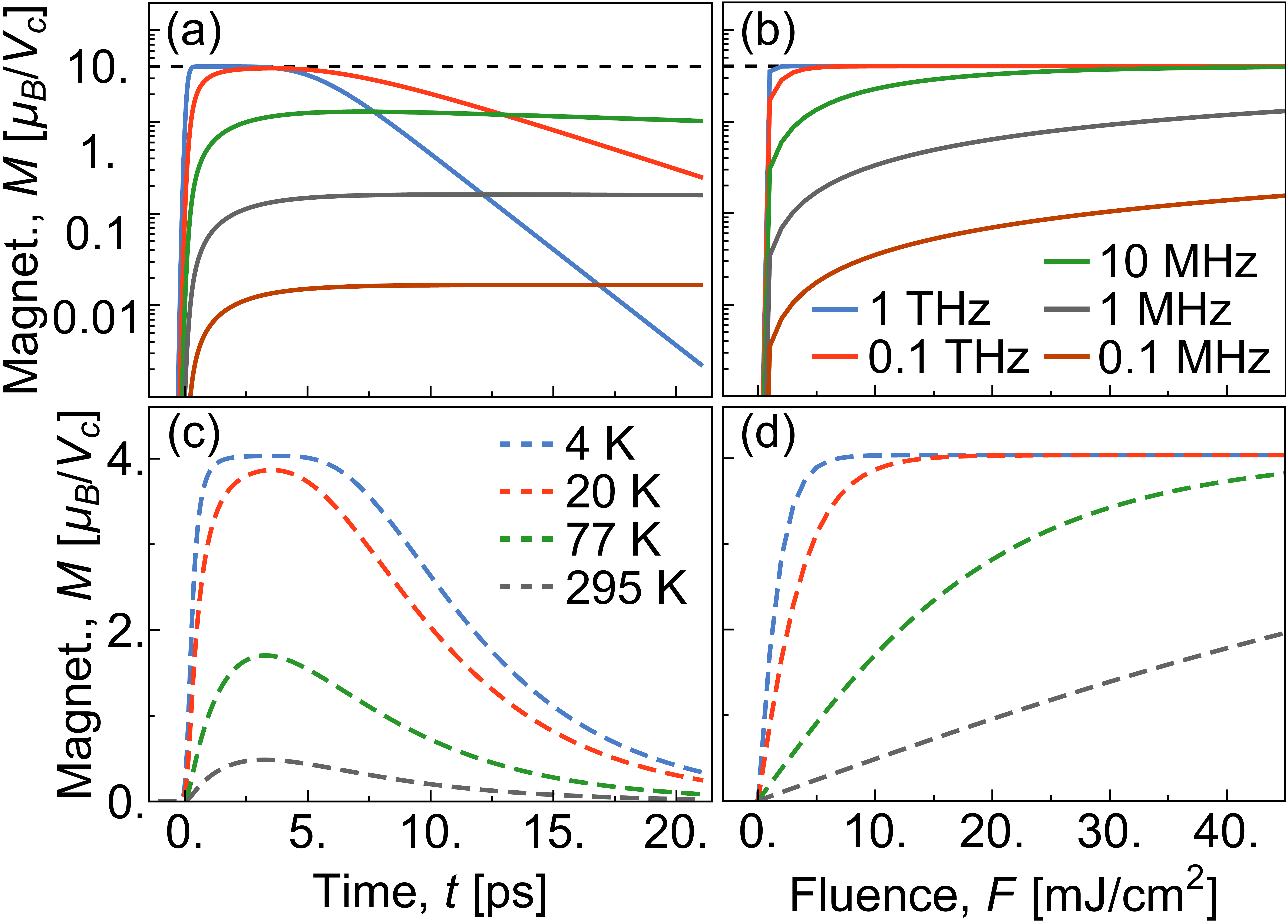}
\caption{
Magnetization, $M=m/V_c$, induced by the $E_{1u}(4.8)$ mode when excited by a circularly polarized terahertz pulse with a duration of 350~fs. (a) Time evolution of $M$, varying with the decay rate, $\gamma_0$, for a fluence of 10~mJ/cm$^2$ at 4~K. The dashed line marks the saturation magnetization. (b) Fluence dependence of $M$, varying with $\gamma_0$. (c) Time evolution of $M$ varying with temperature for a fluence of 10~mJ/cm$^2$ for $\gamma_0=1$~THz. Shown are graphs for the boiling temperatures of helium (4~K), hydrogen (20~K), and nitrogen (77~K), as well as for room temperature (295~K). (d) Fluence dependence of $M$, varying with temperature.}
\label{fig:magnetizationdependence}
\end{figure}

We now vary the strength of the excitation. We show the maximum amplitudes of the effective magnetic fields for a range of experimentally accessible fluences of the terahertz pulse \cite{Liu2017} in Fig.~\ref{fig:dynamics}(c), where we fix the pulse duration at $\tau = 350$~fs. The effective magnetic fields depend linearly on the fluence and reach 11.4~T for the $E_{1u}(5.9)$ mode and 107~T for the $E_{1u}(4.8)$ mode at a fluence of 40~mJ/cm$^2$. In order to ensure experimental feasibility, we evaluate the atomic displacements along the eigenvectors of the phonon modes. The Lindemann stability criterion predicts melting of the crystal lattice when the root mean square displacements reach between 10\%{} and 20\%{} of the interatomic distance \cite{Lindemann1910}. We extract the maximum root mean square displacements as $d=\mathrm{max}_{n} | \mathbf{d}_{n}/\sqrt{2} |$, where $\mathbf{d}_{n} = \mathbf{q}_{n} Q_a(t)/\sqrt{\mathcal{M}_n}$ is the displacement of ion $n$. Even at fluences of 40~mJ/cm$^2$, the largest root mean square displacements of the chloride ions reach only 1.3\%{} of the interatomic distance of $2.97$~\AA{} for the $E_{1u}(5.9)$ mode and 3.8\%{} for the $E_{1u}(4.8)$ mode, well below the vibrational damage threshold. Note that other effects may occur, e.g. Zener tunneling, that are not accounted for here. At these high fields, nonlinear couplings between coherently excited infrared-active modes and other vibrational degrees of freedoms come into play \cite{forst:2011,subedi:2014}. These modes do not contribute directly to the spin-phonon coupling however, and we therefore neglect the effect of nonlinear phonon-phonon coupling in this context. Furthermore, the centrosymmetry of CeCl$_3$ prevents nonlinear optical effects, such as second-harmonic generation, to occur at high fluences.


Next, we look at the magnetization, $M=m/V_c$, that can be induced in CeCl$_3$ according to Eqs.~(\ref{eq:magnetization}) and (\ref{eq:rateequation1}). In Fig.~\ref{fig:magnetizationdependence}, we show the evolution of the magnetization in response to the effective magnetic field generated by the $E_{1u}(4.8)$ mode when excited with a resonant terahertz pulse with a duration of 350~fs and a fluence of 10~mJ/cm$^2$, as well as the dependence of the magnetization on the fluence of the laser pulse. In Figs.~\ref{fig:magnetizationdependence}(a) and \ref{fig:magnetizationdependence}(b), we vary the decay rate, $\gamma_0$, while keeping the temperature fixed at 4~K, and in Figs.~\ref{fig:magnetizationdependence}(c) and \ref{fig:magnetizationdependence}(d) we vary the temperature, while keeping $\gamma_0=0.1$~THz. For fast decay rates and at low temperatures, even small fluences of $<$10~mJ/cm$^2$ are sufficient to fully polarize the spins of the material, yielding a transient saturation magnetization of $M=4~\mu_B/V_c$. 
The slower the decay rate and the higher the temperature, the higher the fluence of the laser pulse has to be in order to induce a significant magnetization. The influence of the decay rate on the achievable magnetization is hereby much larger than that of temperature. The slowest decay rate of $\gamma_0=0.1$~MHz that we look at here corresponds to the nanosecond timescale, on which the thermodynamic picture of spin polarization is known to hold \cite{vanderziel:1965,Pershan1967}. Therefore, the corresponding values of induced magnetization for $\gamma_0=0.1$~MHz in Fig.~\ref{fig:magnetizationdependence}(b) can be regarded as lower boundary.


\section{Discussion}

Our predictions can be experimentally realized in state-of-the-art tabletop setups that provide terahertz pulses in the required frequency range \cite{Liu2017,Vicario2020}, where the phonon-induced magnetization of the material can be probed by Faraday rotation measurements. Tuning the frequency of the terahertz pulse in and out of resonance with the phonon modes can distinguish a possible contribution of the optical inverse Faraday effect to the magnetization 
from the phonon-induced mechanism. The effective magnetic fields calculated here reach 11.4~T for the $E_{1u}(5.9)$ mode and 107~T for the $E_{1u}(4.8)$ mode for terahertz pulses with a fluence of 40~mJ/cm$^2$. We further predict that the subsequent spin polarization of the paramagnetic cerium ions can possibly reach full saturation at low temperatures. These magnetic fields and moments are several orders of magnitude larger than those demonstrated in previous experiments and calculations of the phonon inverse Faraday effect, where fields can be found around micro to milli tesla and magnetic moments around the order of nuclear magnetons \cite{nova:2017,juraschek2:2017,Juraschek2020_3,Geilhufe2021}. Furthermore, the fields we predict are orders of magnitude larger than well-established experiments on the optical inverse Faraday effect, where fields of fractions of tesla have been reported for comparable fluences in the visible spectrum for different insulating and semiconducting systems \cite{kimel:2005,reid:2010}. We display a detailed comparison in Table~\ref{tab:magneticfieldstrengths} in the Appendix.

A direct quantitative comparison of the optical and phonon inverse Faraday effects remains difficult, as no practical and general ab-initio theory of the mechanisms exists to date, and phonon-pumping experiments have only become feasible in very recent years. This comparison is further complicated by the breakdown of the effective magnetic field picture for pulse durations on the order of tens of femtoseconds \cite{reid:2010,popova:2011,Popova2012,Mikhaylovskiy2012}. In the future, explicit calculations of spin-phonon decay rates \cite{Lunghi2019} and coupling strengths will therefore be necessary to further quantify the timescale and magnitude of the optical and phonon inverse Faraday effects and to make predictions for a broad range of materials. First steps have been made over the course of the last decade \cite{popova:2011,Popova2012,battiato:2014,Mondal2015,Majedi2020}, and first quantitative results for the optical inverse Faraday effect have been achieved in metals \cite{Berritta2016,Dannegger2021}.

While we have chosen CeCl$_3$ as our model system, the mechanism described here should be general to the entire class of rare-earth trihalides \cite{schaack:1976,schaack:1977} and possibly to $4f$ magnets in general, as similar magnitudes of the spin-phonon coupling have been found in ferromagnetic LiTbF$_4$ \cite{Dorfler1983} and materials exhibiting the phonon Hall effect, such as paramagnetic Tb$_3$Ga$_5$O$_{12}$ \cite{sheng:2006,zhang:2014}. A future question to answer is whether spin-phonon couplings in $3d$ magnets can reach similar magnitudes to those in $4f$ magnets. Potential giant phonon-induced effective magnetic fields in the paramagnetic phases of $3d$ magnets would directly impact a large variety of materials that are already being used in magnetoelectronic technologies \cite{Bader2010,Spaldin2019}. Another future question is whether coherent chiral phonon excitation could stabilize an ordered spin phase above the equilibrium ordering temperature in paramagnets, similar to phonon- and light-induced superconductivity above the equilibrium critical temperature in superconductors \cite{fausti:2011,mankowsky:2014,Mitrano2016,Cantaluppi2018}.


\begin{acknowledgments}
We are grateful to Christian Tzschaschel, Jannis Lehmann, Shovon Pal, Nicola Spaldin, Michael Fechner, Ankit Disa, Alexander von Hoegen, and Andrea Cavalleri for useful discussions. This project was supported by the Swiss National Science Foundation (SNSF) under Project ID 184259 and the DARPA DRINQS Program under Award No. D18AC00014. P.N. is a Moore Inventor Fellow and gratefully acknowledges support from the Gordon and Betty Moore Foundation through Grant No. GBMF8048. Calculations were performed at the National Energy Research Scientific Computing Center (NERSC), supported by the Office of Science of the U.S. Department of Energy under Contract No. DE-AC02-05CH11231.
\end{acknowledgments}


\begin{appendix}

\section*{Appendix: Phonon-frequency splitting and magnetization in $4f$ paramagnets}

In this section, we provide detailed derivations of the equations used in the \textit{Spin-phonon coupling and coherent phonon dynamics} section of the main text. Motions of the ions along the eigenvectors of doubly degenerate phonon modes modify the crystal electric field (CEF) around the paramagnetic ions and induce virtual transitions between the ground-state energy levels and higher-lying CEF states, see Fig.~(2) in the main text. The spin states of rare-earth ions in compounds are close to those of the free ions and the total angular momentum (isospin), $J$, is a good quantum number. In CeCl$_3$, the lowest energy level has $J=5/2$, which splits into three Kramers doublets, of which $m_J=\pm5/2$ is the ground state. the interaction of chiral phonons with the isospin can be written as an effective ``spin-orbit'' type Hamiltonian \cite{Ray1967,Capellmann1991,Ioselevich1995,sheng:2006,Kagan2008,Wang2009,zhang:2014},
\begin{equation}\label{eq:App_spinphonon}
H^{\mathrm{s-ph}} = \sum\limits_{\alpha n} k_{\alpha n} \mathbf{J}_\alpha \cdot \mathbf{L}_{\alpha n},
\end{equation}
where $\mathbf{J}_\alpha$ is the total isospin of unit cell $\alpha$, $\mathbf{L}_{\alpha n}$ is the phonon angular momentum generated by mode $n$, and $k_{\alpha n}$ is the coupling coefficient. The index $\alpha$ runs over all unit cells of the crystal and $n$ over all chiral phonon modes. 

For optical phonons at the Brillouin-zone center, the phonon angular momentum is homogeneous across unit cells and we can drop the index $\alpha$. It is given by $\mathbf{L}_n=\mathbf{Q}_n\times\dot{\mathbf{Q}}_n$, where $\mathbf{Q}_n=(Q_{na},Q_{nb},0)$ contains the normal mode coordinates of the two orthogonal components of a doubly degenerate phonon mode, $Q_{na}$ and $Q_{nb}$, in the $ab$ plane of the crystal. We can further treat the isospin in a mean-field approximation and replace $\mathbf{J}_\alpha$ by the ensemble average, $\Braket{\mathbf{J}_\alpha}$. Taking into account only isospin components perpendicular to the doubly degenerate phonon modes, the ensemble average of the isospin is $\Braket{\mathbf{J}_\alpha}=2|\mathbf{J}|\mathbf{e}_z(\Braket{n_{-J}}-\Braket{n_J})$, where $\mathbf{e}_z$ is a unit vector along the $c$ axis of the crystal. $\mathbf{J}$ is the isospin of a single cerium ion, of which there are two per unit cell. $\Braket{n_{\pm J}}$ is the Fermi-Dirac distribution describing the occupation of the ground-state Kramers doublet. The magnetic moment per unit cell, $\mathbf{m}$, is then given by
\begin{equation}\label{eq:App_magnetization}
\mathbf{m} = g_J \mu_B \Braket{\mathbf{J}_n}=2 g_J \mu_B \sqrt{J(J+1)} \mathbf{e}_z \left(\Braket{n_{-J}}-\Braket{n_J}\right),
\end{equation}
where $g_J$ is the Land\'{e} factor. Eq.~(\ref{eq:App_magnetization}) is the same as (2) in the main text. The theoretical value of the prefactor in Eq.~(\ref{eq:App_magnetization}) for the $m_J=\pm5/2$ ground-state doublet is $g_{\pm5/2} = g_{J} \sqrt{J(J+1)} = 2.54$, which is reasonably close to the experimental value of 2.02 \cite{Thalmeier1977}, showing that the orbital angular momentum is mostly unquenched. We can now rewrite Eq.~(\ref{eq:App_spinphonon}) in terms of the magnetic moment, yielding
\begin{equation}\label{eq:App_newspinphonon}
H^{\mathrm{s-ph}} = \sum\limits_{n} K_{n} \mathbf{m} \cdot \mathbf{L}_n,
\end{equation}
where we redefined the coupling as $K_n \mathbf{m} = k_{\alpha n}\Braket{\mathbf{J}_\alpha}$. If we look at the interaction one mode at a time, we can drop the index $n$, which yields Eq.~(1) from the main text.

The phonon Lagrangian, $\mathcal{L}$, including the interaction in Eq.~(\ref{eq:App_newspinphonon}), can be written as
\begin{eqnarray}\label{eq:App_Lagrangian}
\mathcal{L}(Q,\dot{Q}) & = & \frac{1}{2} \dot{Q}_a^2 + \frac{1}{2} \dot{Q}_b^2 - \frac{\Omega^2_0}{2} Q_a^2 - \frac{\Omega^2_0}{2} Q_b^2 \nonumber\\ & & - K m (Q_a \dot{Q}_b - Q_b \dot{Q}_a),
\end{eqnarray}
where 
$\Omega_a=\Omega_b\equiv\Omega_0$ is the eigenfrequency of the doubly degenerate phonon mode and $m=|\mathbf{m}|$. In frequency space, the Lagrangian can be transformed according to $Q_a \rightarrow Q_{\Omega a}\exp(i\Omega t)$, $\dot{Q}_a \rightarrow i\Omega Q_{\Omega a}\exp(i\Omega t)$, and $Q_a \dot{Q}_b \rightarrow i\Omega(Q_{\Omega a}^* Q_{\Omega b} - Q_{\Omega a} Q_{\Omega b}^*)/2$ to yield
\begin{equation}\label{eq:App_frequencyspace}
\mathcal{L}(Q_\Omega,Q_\Omega^*) = \mathbf{Q}_\Omega \mathbf{D} \mathbf{Q}_\Omega^*,
\end{equation}
where $\mathbf{Q}_\Omega=(Q_{\Omega a},Q_{\Omega b},0)$, and $\mathbf{D}$ is the dynamical matrix. The spin-phonon coupling modifies the dynamical matrix as $\mathbf{D}(m) = \mathbf{D}^{(0)} + i\mathbf{D}^{(1)}(m)$ \cite{anastassakis:1972,Holz1972,schaack:1976,schaack:1977,dzyaloshinskii:2009,riseborough:2010,juraschek2:2017,Liu2017_phonondichroism,Juraschek2020_3,Baydin2022}. In order to evaluate the effect of the interaction on the phonon frequencies, we compute the determinant of the dynamical matrix, which contains the spin-phonon coupling in its off-diagonal components,
\begin{equation}\label{eq:App_determinant}
\mathrm{det}~\mathbf{D} = 
\left| \begin{array}{cc} \Omega^2-\Omega^2_0 & -2i\Omega K m  \\ 
2i\Omega K m & \Omega^2-\Omega^2_0 \end{array} \right|.
\end{equation}
Solving the determinant close to the Brillouin-zone center ($\Omega\rightarrow\Omega_0$) yields Eq.~(3) from the main text, describing a splitting of the frequencies of right- and left-handed circular polarizations, $\Omega_\pm$, of the doubly degenerate phonon mode,
\begin{equation}\label{eq:App_phononsplittinglinear}
\Omega_\pm(m) = \Omega_0\sqrt{1\pm\frac{2 K m}{\Omega_0}} \approx \Omega_0 \pm K m.
\end{equation}

Without an external magnetic field, the energy levels of the $m_J=\pm 5/2$ ground-state Kramers doublet are degenerate, there is no net magnetic moment per unit cell, and the phonon frequencies in Eq.~(\ref{eq:App_phononsplittinglinear}) remain degenerate. Applying a magnetic field, $B\parallel c$, to the paramagnet splits the ground-state doublet, $\Delta E = E_{-5/2}-E_{5/2} = 2 g_{\pm5/2} \mu_B B$, and induces a magnetization given by
\begin{eqnarray}\label{eq:App_inducedmagnetization}
m & = &  2g_{\pm5/2}\mu_B \left(\Braket{n_{-5/2}} - \Braket{n_{5/2}}\right) \nonumber\\
& = & 2g_{\pm5/2} \mu_B \Big[\frac{1}{\exp\left(-\frac{g_{\pm5/2} \mu_B B}{k_B T}\right)+1} \nonumber \\ & & \phantom{XXXXX} - \frac{1}{\exp\left(\frac{g_{\pm5/2} \mu_B B}{k_B T}\right)+1}\Big] \nonumber \\
& = & 2g_{\pm5/2}\mu_B \tanh\left(\frac{g_{\pm5/2} \mu_B B}{2k_B T}\right).
\end{eqnarray}
In the above equation, we have used the relation
\begin{equation}
\frac{1}{\exp(-x)+1}-\frac{1}{\exp(x)+1}=\tanh\left(\frac{x}{2}\right).
\end{equation}
Inserting Eq.~(\ref{eq:App_inducedmagnetization}) into (\ref{eq:App_phononsplittinglinear}) yields Eq.~(4) of the main text,
\begin{eqnarray}\label{eq:App_phononsplittingfull}
\Delta\Omega(B) & = & 2Km = 4 K g_{\pm5/2}\mu_B \tanh\left(\frac{g_{\pm5/2} \mu_B B}{2k_B T}\right) \\
& \equiv & \Delta\Omega_s \tanh\left(\frac{g_{\pm5/2} \mu_B B}{2k_B T}\right),
\end{eqnarray}
where $\Delta\Omega_s$ is the saturation phonon-frequency splitting between right- and left-handed chiral phonon modes. This equation allows us to extract the spin-phonon coupling from experimentally measured phonon splittings, $K=\Delta\Omega_s/(4 g_{\pm5/2}\mu_B)$.


\onecolumngrid

\begin{table*}[h]
\centering
\bgroup
\def\arraystretch{1.5}
\caption{
Comparison of effective magnetic fields ($B_\mathrm{eff}$) induced by the optical and phonon inverse Faraday effects (IFE) in different materials. We further display the pulse fluences, durations, and spectral ranges. Previous theoretical work has most of the time displayed induced magnetizations arising from phonon orbital magnetic moments in terms of nuclear magnetons per unit cell. We convert the effective magnetic field as $B=\mu_0 m/V_c$, where $\mu_0$ is the vacuum permeability and $m$ is the magnetic moment per unit cell. Note that direct quantitative comparisons between the mechanisms are complicated by the fact that the effective magnetic field picture breaks down for very short (tens of femtoseconds) pulses.
}
\begin{tabular}{lllllll}
\hline\hline
Reference & Excitation effect & Material (magnetic state) & $B_\mathrm{eff}$ & Pulse fluence & Pulse duration & Spectral range \\
\hline
Experiment \cite{kimel:2005} & Optical IFE & DyFeO$_3$ (weak FM) & 0.3~T & 30~mJ/cm$^2$ & 200~fs & 800~nm \\
Experiment \cite{reid:2010} & Optical IFE & Dy$_3$Al$_5$O$_{12}$ (PM) & 0.3~T & 200~mJ/cm$^2$ & 50~fs & 800~nm \\
Experiment \cite{nova:2017} & Phonon IFE & ErFeO$_3$ (weak FM) & 36~mT & 20~mJ/cm$^2$ & 130~fs & THz/MIR \\
Theory \cite{juraschek2:2017,Juraschek2019} & Phonon IFE & Various, largest value shown & 0.4~mT& 1.5~mJ/cm$^2$ & 1.5~ps & THz/MIR\\
Theory \cite{Juraschek2020_3} & Phonon IFE & NiO (AFM) & 2.5~mT & 80~mJ/cm$^2$ & 2.25~ps & THz/MIR\\
Theory \cite{Geilhufe2021} & Phonon IFE & KTaO$_3$ (NM) & 5~$\mu$T& $\approx$3~mJ/cm$^2$ & $\approx$2~ps& THz\\
This work & Phonon IFE & CeCl$_3$ (PM) & 107~T & 40~mJ/cm$^2$ & 350~fs &  THz \\
\hline\hline
\end{tabular}
\label{tab:magneticfieldstrengths}
\egroup
\end{table*}

\end{appendix}

\twocolumngrid



%

\end{document}